\documentclass[11pt,a4paper]{article}
\pdfoutput=1

\usepackage{jcappub}

\usepackage{bbm,amsthm,amsmath,amssymb,cancel,multirow,graphics,braket,graphicx,mathtools,calc,ulem,pbox,breakcites,environ,tabularx,hyperref}

\newcommand{\mcl}[1]{\mathcal{#1}}

\newcommand{\sO}[0]{\mathcal{O}}

\newcommand{\sci}[1]{\times10^{#1}}

\DeclareMathOperator{\im}{Im}

\newcommand{\mnras}[0]{Mon.Not.Roy.Astron.Soc.}
\newcommand{\apj}[0]{Astrophys.J.}
\newcommand{\prd}[0]{Phys.Rev.D}

\newcommand{\ssr}[0]{Space Sci.Rev.}
\newcommand{\apss}[0]{Astrophys.Space Sci.}
\newcommand{\aap}[0]{Astronomy \& Astrophysics}
\newcommand{\jcap}[0]{JCAP}

\newcommand{\mbf}[1]{\mathbf{#1}}
\newcommand{\tB}[0]{\tilde{\mathbf{B}}}
\newcommand{\tb}[0]{\tilde{B}}
\newcommand{\bx}[0]{\mathbf{x}}
\newcommand{\bk}[0]{\mathbf{k}}
\newcommand{\bB}[0]{\mathbf{B}}

\newcommand{\bbzet}[0]{\braket{\mathbf{B}\cdot\mathbf{B}\,\zeta}}
\newcommand{\ck}{{\check{k}}}
\newcommand{\bmu}[0]{\tilde B_\mu}

\begin{document}

\title{Probing correlations of early magnetic fields using $\mu$-distortion}

\author{Jonathan Ganc,}
\author{Martin S. Sloth}
\affiliation{CP$^3$-Origins, Center for Cosmology and Particle Physics Phenomenology\\
University of Southern Denmark, Campusvej 55, 5230 Odense M, Denmark}

\emailAdd{ganc@cp3.dias.sdu.dk}
\emailAdd{sloth@cp3.dias.sdu.dk}

\abstract{The damping of a non-uniform magnetic field between the redshifts of about $10^4$ and $10^6$ injects energy into the photon-baryon plasma and causes the CMB to deviate from a perfect blackbody spectrum, producing a so-called $\mu$-distortion. We can calculate the correlation $\langle\mu T\rangle$ of this distortion with the temperature anisotropy $T$ of the CMB to search for a correlation $\langle B^2\zeta\rangle$ between the magnetic field $B$ and the curvature perturbation $\zeta$; knowing the $\langle B^2\zeta\rangle$ correlation would help us distinguish between different models of magnetogenesis. Since the perturbations which produce the $\mu$-distortion will be much smaller scale than the relevant density perturbations, the observation of this correlation is sensitive to the squeezed limit of $\langle B^2\zeta\rangle$, which is naturally parameterized by $b_{\text{NL}}$ (a parameter defined analogously to $f_{\text{NL}}$). We find that a PIXIE-like CMB experiments has a signal to noise $S/N\approx 1.0 \times b_{\text{NL}} (\tilde B_\mu/10\text{ nG})^2$, where $\tilde B_\mu$ is the magnetic field's strength on $\mu$-distortion scales normalized to today's redshift; thus, a 10 nG field would be detectable with $b_{\text{NL}}=\mathcal{O}(1)$. However, if the field is of inflationary origin, we generically expect it to be accompanied by a curvature bispectrum $\langle\zeta^3\rangle$ induced by the magnetic field. For sufficiently small magnetic fields, the signal $\langle B^2 \zeta\rangle$ will dominate, but for $\tilde B_\mu\gtrsim 1$ nG, one would have to consider the specifics of the inflationary magnetogenesis model.

We also discuss the potential post-magnetogenesis sources of a $\langle B^2\zeta\rangle$ correlation and explain why there will be no contribution from the evolution of the magnetic field in response to the curvature perturbation.}

\keywords{cosmic magnetic fields theory, CMBR theory, non-Gaussianity, primordial magnetic fields}

\maketitle
\flushbottom

\section{Introduction}
\label{sec:introduction}

The $\Lambda$CDM model of cosmology provides a very good fit to many cosmological observables. However, there remain many questions about the evolution of the early universe and the forces that shaped it. Over the last decade, much of our cosmological information has come from exploring the temperature perturbations of the mostly uniform Cosmic Microwave Background (CMB) by experiments such as WMAP, Planck, SPT, and ACT. However, at this point, we have nearly exhausted the information in this signal (at least from pre-recombination sources), and future discoveries must come from other sources, e.g. large scale structure, improved standard candle measurements, the 21 cm line, etc. However, the temperature perturbation is not the only information in the CMB, as excitingly suggested by the possible recent $B$-mode detection of BICEP2 \cite{Ade:2014xna}, with other experiments soon to come with more polarization data, including Planck, Spider, the Keck Array, and POLARBEAR. In this paper, we look at still another source of data in the CMB: distortion from a perfect blackbody spectrum, in particular $\mu$-type distortion.

To understand $\mu$-distortion, one must first realize the limits of a widely known fact: the CMB has no measured pre-recombination deviations from a blackbody spectrum (\cite{Fixsen:1996nj,Mather:1993ij,Mather:1991pc}). However, even at recombination, the photon distribution is actually expected to be only an imperfect Planck spectrum \cite{Weymann:1966ab,Zeldovich:1969ab,Sunyaev:1970aq}. In an ideal photon gas, both the energy distribution and number density of photons are specified by the photon temperature $T$; thus, adding energy or entropy to the gas generically requires both a redistribution of photon energy and a change in photon number in order to re-reach a Planck spectrum at some new temperature $T'$. Well before recombination, at $z\gg z_\mu^i =2\sci{6}$, the photon spectrum in the early universe plasma is Planckian because of the efficiency of the various relevant reactions. However for $z\lesssim z_\mu^i$, the primary photon non-conserving process, double-Compton scattering ($\gamma+e^-\to 2\gamma+e^-$), becomes inefficient, while elastic Compton scattering remains efficient; thus, the photon gas responds to entropy increases by reaching thermal equilibrium but with a conserved photon number, which is thermodynamically equivalent to giving the photons a chemical potential $\mu$. The result is that the photon develop a Bose-Einstein distribution given by the appropriate values of $T$ and $\mu$; this deviation is called $\mu$-distortion. Later at $z\lesssim z_\mu^f= 5\sci{4}$, even elastic Compton scattering is inefficient and new perturbations to the photon distribution are minimally reprocessed, usually producing so-called $y$-distortions.

$\mu$-distortion is typically parameterized as a dimensionless number in the photon distribution function $n(x) = (e^{x+\mu}-1)^{-1}$, where $x$ is the dimensionless frequency $x\equiv h\nu/k_B T$. COBE/FIRAS constrained $|\mu|< 9\sci{-5}$ \cite{Fixsen:1996nj}, which was marginally improved by the TRIS experiment to $|\mu|< 6 \sci{-5}$ \cite{Zannoni:2008xx,Gervasi:2008eb}. The proposed PIXIE experiment would considerably improve the bound to $|\mu|\lesssim 9\sci{-8}$ \cite{Kogut:2011xw}. 

$\mu$-distortion is produced by effects that inject energy/entropy into the photon-baryon plasma during the $\mu$-distortion era $z_\mu^f < z < z_\mu^i$. There is one inevitable source of such injections, namely Silk damping of density perturbations \cite{Silk:1968,Peebles-Yu:1970,Kaiser:1983}. Particle decay during the appropriate era can also produce $\mu$-distortion \cite{Hu-Silk-Relic-pcls:1993}, though the decay time must be fine-tuned and one must take care to satisfy constraints on the radiation energy density from BBN and the CMB. In this paper, we will focus on the production of $\mu$-distortion from the decay of non-uniform magnetic fields due the viscosity of the photon-baryon plasma \cite{Jedamzik:1999bm}.

Most $\mu$-distortion studies have focused on the monopole $|\mu|$, i.e. the sky-averaged $\mu$-distortion. Recently, \cite{Pajer&Zald:2012-New-window}\footnote{See also \cite{Ganc:2012ae}.} looked at the signatures of anisotropic $\mu$-distortion. In their case, they considered the $\mu$-distortion caused by the damping of the density perturbation $\zeta$. Since the entropy for the $\mu$-distortion comes from spectral power (or equivalently, from the energy in the density waves), one has that $\mu\propto \zeta^2$. They then correlated the $\mu$-distortion signal with the larger temperature anisotropy signal $T$, which is primarily given by $T\propto\zeta$. Thus, the $\braket{\mu T}$ correlation constrains the squeezed limit of the bispectrum $\braket{\zeta^3}$, i.e. $f_{\text{NL}}^\text{loc}$. 

As noted earlier, the damping of magnetic fields also produce $\mu$-distortion \cite{Jedamzik:1996wp,Jedamzik:1999bm}, proportional to the damped energy $B^2$. It is probable that there are $\sim\mu$G magnetic fields in galaxies and clusters (e.g. \cite{Bonafede:2010xg,Beck:2012ag}), and their origins are still speculative. More recently, there have been tantalizing hints of large-scale magnetic fields in the intergalactic medium with strengths of $\mcl{O}(10^{-15} - 10^{-20})$ G \cite{Tavecchio:2010mk,Neronov:1900zz,Takahashi:2013uoa} and which are incompatible with most magnetogenesis scenarios (e.g. from the QCD or electroweak phase-transition). Inflation can, in principle, produce such large-scale fields, though taking the BICEP2 data \cite{Ade:2014xna} at face values strongly disfavors inflationary production as well  \cite{Kandus:2010nw,Ferreira:2013sqa,Fujita:2014sna,Ferreira:2014af}. 

Just as with the curvature perturbation, we can study cosmological magnetic fields through their correlations and indeed, to study magnetic fields, it makes sense to consider a possible correlation $\bbzet$ with the already-detectable curvature perturbation signal. In analogy with $f_{\text{NL}}$, we can suppose that a correlated magnetic field $\bB^\text{corr}$ is produced from an uncorrelated field $\bB^\text{uncor}$ \cite{Jain:2012ga} as
\begin{align}
  \bB^\text{corr} \simeq \bB^\text{uncor} + b_{\text{NL}} \bB^\text{uncor} \zeta + \ldots\,,
\end{align}
so that
\begin{align}\label{eq:dfnn-of-bnl}
  \left\langle\bB^*(\bk_1) \cdot \bB(\bk_2)\, \zeta(\bk_3) \right\rangle_{k_3\ll k_1\approx k_2} = (2\pi)^3 b_{\text{NL}} 
    \delta^{(3)}\!(-\bk_1 + \bk_2 + \bk_3) P_B(k_1) P_\zeta(k_3)
\end{align}
in the squeezed limit $k_3\ll k_1\approx k_2$. Actually, we take (\ref{eq:dfnn-of-bnl}) to be the definition of $b_{\text{NL}}$, with the understanding that we are primarily interested in the squeezed limit. Generically, one expects that inflationary magnetogenesis produces\footnote{A consistency relation for a simple class of models was derived in \cite{Jain:2012ga,Jain:2012vm}, yielding $b_{NL}=(n_B-4)$.} $|b_{\text{NL}}|\gtrsim 1$. The $\braket{\mu T}$ correlation first explored by \cite{Pajer&Zald:2012-A-hydrodynamical} offers a promising way to probe such a correlation, which we explore here. Related ideas of how to probe $b_{NL}$ with large-scale structure consistency relations was put forward in \cite{Berger:2014wta}. 

In this work, we demonstrate that a primordial $\bbzet$ correlation\footnote{Note that, by ``primordial'', we mean existing in the era of magnetogenesis, not (as is sometimes meant) merely happening anytime before recombination.} could potentially be observable in a measurement of the $\braket{\mu T}$ correlation if $b_{\text{NL}}=\sO(1)$ and the magnetic field on $\mu$ scales is $\sO(10\text{ nG})$. However, one also needs to consider that magnetic field energy density is inherently non-Gaussian (since it goes as $B^2$) so that magnetic fields inevitably source some level of the three-point function $\langle\zeta^3\rangle$, which also generates a $\braket{\mu T}$ signal. If the magnetic fields were of inflationary origin, there would be sufficient time for this correlation to grow large. For sufficiently small magnetic fields, the $\bbzet$ signal will still dominate because it goes as $B^2$ whereas other signals go as higher powers of the magnetic field. However, for $B_\mu\gtrsim$ 1 nG, one has to consider the specifics of the inflationary magnetogenesis model to determine which signal would be detected first.

Our work is organized as follows. In Sec. \ref{sec:measuring-mgntc-flds-mu-distortion}, we discuss how magnetic fields damp in the early universe, producing $\mu$-distortion; we then correlate this with the temperature perturbation $T$ to find a fairly generic formula for $C_l^{\mu T}$ in terms of $\bbzet$ without regard to the source of the correlation. In the next section, Sec. \ref{sec:primordial-bnl}, we quickly derive the formula for the signal from a primordial $b_{\text{NL}}$. In the next two sections, we consider two other sources for a $\bbzet$ correlation. In Sec. \ref{sec:evln-of-zet-due-to-B}, we repeat \cite{Miyamoto:2013oua}'s recent calculation of the correlation produced by the evolution of $\zeta$ due to the anisotropic stress of the magnetic field; they found (and we confirm) a very small signal. In Sec. \ref{sec:evln-of-B-due-to-zet}, we consider the reverse effect, the potential correlation caused by the evolution of $B$ due to the initial adiabatic perturbation, finding also no relevant impact. We discuss the observational impact of our results in Sec. \ref{sec:observ-prim-BBzet}, including the consequences of the competing $f_{\text{NL}}$ signal, and then conclude in Sec. \ref{sec:conclusion}.

\subsection{Notation and conventions}
\label{sec:notation}

We use the following conventions:
\begin{itemize}
\item We use $t$ for physical time and $\eta$ for conformal time; dots denote derivatives with respect to $t$ and primes denote derivatives with respect to $\eta$.

\item $f_{\text{NL}}$ refers to $f_{\text{NL}}^\text{loc}$. 

\item $\rho_r$ refers to the total density in relativistic particles, which we calculate assuming that $n_\text{eff}=3.046$ and that neutrinos were relativistic through recombination. 

\item We use as $\zeta$ the same variable that \cite{Weinberg:2008zzc} calls $\mcl{R}$. Note that, in comparison with \cite{Shaw:2009nf}, $\zeta = -\zeta^\text{SL}$. 

\item Given that the Hubble expansion causes magnetic fields to decay as $B\propto a^{-2}$, we will often use a ``comoving'' magnetic field
\begin{align}
  \tB \equiv B a^2\,,
\end{align}
and comoving energy density
\begin{align}
  \tilde\rho \equiv \rho a^4\,,
\end{align}
so that $\tb$, $\tilde\rho$ are essentially normalized to their present values (e.g. $\tilde\rho_\gamma\approx\rho_{\gamma0}$). 

\item When we omit an explicit time dependence for a magnetic field, e.g. $\tB(\bk)$, we take it to be the field strength before any damping on relevant scales.

\item We relate the two point function $\braket{\tb_i^*\tb_j}$ and $P_B$ via
\begin{align}\label{eq:B-2_pt_fcn}
  \Braket{\tb_i^*(\bk_1)\tb_j(\bk_2)} &=
   (2\pi)^3 \frac{P_{ij}(\hat k_1)}{2} P_B(k_1) \delta(\bk_1 - \bk_2)\,;
   &&\text{where}
   & P_{ij}(\hat k) &=\delta_{ij}-\hat{k}_i\hat{k}_j\,,
\end{align}
we have neglected a helical component which we will not consider here. 

\item We refer to the damping scale at the start, finish of the $\mu$-distortion era as $k_D^i = 2.1\times10^{4}\text{ Mpc}^{-1}$, $k_D^f = 83 \text{ Mpc}^{-1}\,,$. We also define the power damping scale $\ck_D \equiv k_D/\sqrt{2}$. (See Sec. \ref{sec:damp-mgntc-flds} for more information).

\item When performing calculations, we use values from Planck \cite{Planck-overview}.

\item We assume that, besides the primordial magnetic field, the primordial perturbations are adiabatic.
\end{itemize}

\section{Measuring magnetic fields through $\mu$-distortion}
\label{sec:measuring-mgntc-flds-mu-distortion}

\subsection{Damping of magnetic fields in the early universe}
\label{sec:damp-mgntc-flds}

We are interested in determining the energy lost from magnetic fields due to damping, since this energy then goes into producing $\mu$-distortion. We do not expect any energy contribution from electric fields since, with its high concentration of ions, the plasma is a near perfect conductor and quickly cancels out any electric fields at a very early time \cite{Durrer:2013pga}. The evolution of magnetic fields is considerably more complicated. The photons interact with the electrons which interact with the baryons, and the latter two both interact with any magnetic fields \cite{Jedamzik:1996wp}. Since the matter and photons behave together as a fluid, this is properly described as a magnetohydrodynamic (MHD) system. Such a system displays the complexity of standard fluid dynamics, including regimes of turbulence, as well as new phenomenon like additional types of waves beyond standard acoustic waves. In the presence of a background magnetic field, we can identify two new types of waves:

\begin{itemize}
\item \textit{Alfv\'en waves} propagate generally along the background field, while the magnetic field and fluid velocity oscillations are perpendicular to the background field (and the propagation direction). They travel at the speed $v_A \cos\theta$, where, for weak background fields (i.e. $\rho_B\ll\rho_r$), the Alfv\'en velocity is
  \begin{align}
    v_A^2 = \frac{3}{2} \frac{\rho_B}{\rho_r}\,,
  \end{align}
$\rho_B$ is the energy of the background magnetic field, and $\rho_r$ is the energy in relativistic particles. Alfv\'en waves do not involve density fluctuations.

\item Magnetosonic waves propagate at an arbitrary angle with respect to the background field and involve plasma velocity oscillations both parallel and perpendicular to the background field. Magnetosonic waves can be further divided into \textit{fast magnetosonic waves} and \textit{slow magnetosonic waves}. For $\rho_B\ll\rho_r$, the latter travel at the Alfv\'en velocity and also have close to vanishing density fluctuations, while the former travel at nearly the speed of sound and do have density fluctuations.
\end{itemize}

We focus exclusively on Alfv\'en and slow magnetosonic waves, which for our purposes have identical properties. Many papers \cite{Jedamzik:1996wp,Subramanian:1997gi,Jedamzik:1999bm,Banerjee:2004df} have argued that the primordial plasma is nearly incompressible (i.e. $\partial_t \tilde\rho\equiv0$) because the fluid velocity is small compared to the speed of sound -- a sufficient condition at least in classical steady-flow fluids -- or because the magnetic energy density is small. However, \cite{Durrer:2013pga} respond that we do not in fact know the fluid velocity or characteristic time scales on the small scales relevant for magnetic fields in the early universe. Here, we simply assume all waves are Alfv\'en and slow magnetosonic; at worst, this should introduce a $\mcl{O}(1)$ difference to our results.

In \cite{Jedamzik:1996wp} the damping of magnetic fields in a photon-baryon plasma was considered assuming a small uniform background field ($\rho_{B}\ll \rho_r$) with even smaller magnetic perturbations, and then describing the damping of these perturbations. While this scenario does not precisely describe the early universe, which would not have had a clean split between the background and the perturbations, it seems a fair approximation: at a given time, the stochastic variations in superhorizon magnetic fields (if present) would appear as a roughly uniform magnetic background. Also, \cite{Subramanian:1997gi} calculated the magnetic field evolution in special cases where they did not need a linear approximation (i.e., where they did not assume a clean divide between a background magnetic field and a much smaller perturbation) and found results that corroborated those from \cite{Jedamzik:1996wp}.

\cite{Jedamzik:1996wp,Jedamzik:1999bm} use a WKB approximation for the evolution of the magnetic fields
\begin{align}
  \tB(\bk,t) = \tB(\bk) \exp\Big(i\int_{t_I}^t \omega dt\Big)\,,
\end{align}
where $t_I$ is some earlier time, before any damping on the scales we are considering, and $\tB$ is defined in Sec. \ref{sec:notation}. As magnetic field modes enter the horizon during the $\mu$-era, they start off in the photon diffusion limit, where the mode's wavelength $1/k$ is longer than the photon mean free path $l_\gamma \approx 1.96\sci{6}\text{ Mpc}\times a^3$ (for $z\gg1000$),
as well as in the oscillatory regime, where the magnetic fields and velocities oscillate frequently as the mode damps. The dispersion relation here for both Alfv\'en and slow magnetosonic waves is given by 
\begin{align} \label{eq:omega-diff-osc-regime}
  \omega^{\text{SM,A}}_k = v_A \cos\theta \Big(\frac{k}{a}\Big) 
    + i \frac{1}{10} \left(\frac{\rho_\gamma}{\rho_r}\right) l_\gamma \Big(\frac{k}{a}\Big)^2\,,
\end{align}
where $\rho_r$ is the radiation energy density so that, assuming three massless neutrinos decoupling before electron-positron annihilation, we get $\rho_\gamma/\rho_r = \left(1 + 3\left(\frac{7}{8}\right) \left(\frac{4}{11}\right)^{4/3}\right)^{-1}$ \cite{Weinberg:2008zzc}.

The damping scale $k_D$ can be found (neglecting the real part of $\omega$, which does not contribute)
\begin{align}\label{eq:mgntc-damping}
  \exp\Big(i\int_{t_I}^t \omega dt\Big)
    &\approx \exp\left(-\int_{t_I}^t\im\omega_k dt\right) \approx \exp\left( - k^2/k_D^2(t) \right)\,,
\end{align}
where (for $z\gg1000$) $k_{D}(a) = 7.44\times10^{-6}\text{ Mpc}^{-1} \times a^{-3/2}$. Thus, we find that
\begin{align}
  k_D^i &= 2.1\times10^{4}\text{ Mpc}^{-1}\,,
    & k_D^f &= 83 \text{ Mpc}^{-1}\,,
\end{align}
where $k_D^i$, $k_D^f$ are the damping scales at the start, finish of the $\mu$-distortion era. Since we are interested in the damping of power $\tB_{\bk_1}\cdot\tB_{\bk_2}\propto\exp\big(-(k_1^2 + k_2^2)/k_d^2\big)$ in the limit $k_1\approx k_2$, we also find it useful to define the power damping scale 
\begin{align}\label{eq:pow-damping-scale}
  \ck_D \equiv \frac{k_D}{\sqrt{2}}\,.
\end{align}

For comparison, the photon diffusion scale (aka Silk damping scale) $k_{D\gamma}$ \cite{Kaiser:1983} is
\begin{align}\label{eq:photon-diff-scale}
  k_{D\gamma} \approx 4.1\sci{-6}\text{ Mpc}^{-1} a^{3/2}\,;
\end{align}
the photon diffusion scale and the magnetic damping scale are nearly the same, which is not surprising since both proceed via viscosity from photon diffusion.

Let us briefly comment on some other papers that also considered $\mu$-distortion produced by the damping of magnetic fields. While we calculate the damping scale in the same way as \cite{Jedamzik:1999bm}, our method differs from those in \cite{Kunze:2013uja} and \cite{Miyamoto:2013oua}. \cite{Kunze:2013uja} consider damping in the free-streaming limit (where the mode wavelength $1/k\ll l_\gamma$) and overdamped regime; their damping takes place on smaller scales and is applicable if the damping we consider is not efficient for some reason. \cite{Miyamoto:2013oua} adopt a more complicated method for determining the damping scale, where they try to synthesis the results from the different damping regimes specified in \cite{Jedamzik:1996wp}, though notably, the actual damping scales this gives are quite similar to ours. While an interesting idea, there are two reasons why this approach is not an improvement. One, all of the results in \cite{Jedamzik:1996wp} are in particular limits and are not applicable over the intermediate scales connecting the different regimes. Two, \cite{Jedamzik:1996wp} is a linear approximation to the full, potentially complicated MHD behavior and likely does not give sufficiently precise results for a very precise calculation of the damping scale.

\subsection{Calculating $\mu$-distortion and finding $C_l^{\mu T}$}
\label{sec:prod-mu-dist}

$\mu$-distortion has experienced a resurgence of interest in recent years, with a number of papers studying it, e.g. \cite{Khatri:2011aj,Khatri:2012rt,Khatri:2012tw,Chluba:2012gq,Pajer&Zald:2012-A-hydrodynamical,Pajer&Zald:2012-New-window,Ganc:2012ae,Kunze:2013uja,Miyamoto:2013oua}.
It results from the injection of energy $\Delta\rho$ into the photon-baryon plasma during the $\mu$-era ($z_\mu^f = 5\sci{4} < z < z_\mu^i = 2\sci{6}$) and is calculated by \cite{Hu:1992dc,Pajer&Zald:2012-A-hydrodynamical}
\begin{align}\label{eq:mu-distortion}
  \mu \simeq 1.4 \frac{\Delta\rho_\gamma}{\rho_\gamma}\,, \qquad \text{for } \mu\ll1\,,
\end{align}
where $\rho_\gamma$ is the photon energy which is also roughly the plasma energy.

To find the observable signature of $\mu$-distortion, we need to project it onto the 2-dimensional surface that is observed in the sky, i.e. decompose it in terms of the spherical harmonics $Y_{lm}(\bx)$
\begin{align}
  \mu(\bx) = \sum_{lm} a_{lm}^\mu Y_{lm}(\bx)\,.
\end{align}
The energy in a magnetic field (accounting for dissipation) is
\begin{align}
  \tilde\rho_B(\bx,t) &= \frac{\tB(\bx,t)\cdot\tB(\bx,t)}{2\mu_0}
   = 2 \mu_0^{-1} \int \frac{d^3 k_1 d^3 k_2}{(2\pi)^6}
    \tB(\bk_1) \cdot\tB^*(\bk_2) 
    e^{-(k_1^2 + k_2^2)/k_D^2(t)}
    e^{i(\bk_1-\bk_2)\cdot\bx}\,;
\end{align}
as noted in Sec. \ref{sec:notation}, if we omit the time on a magnetic field term, e.g. $\tB(\bk_1)$, we take it to refer to the magnetic field strength before damping. Then we can use the planar wave decomposition
\begin{align}
  e^{i\bk \cdot \bx}
   &=4\pi\sum_{l,m}i^{l}Y_{lm}(\hat{\mbf n})Y^{\ast}_{lm}(\hat\bk)j_{l}(r_L k)
\end{align}
to find
\begin{align}\label{eq:almu}
  a^\mu_{lm} &= \int d^2\hat n \mu_B(\bx) Y_{lm}^*(\hat{\mbf n})
    \approx \frac{1.4}{\rho_{\gamma0}} \int d^2\hat n \left[\tilde\rho_B(\bx,z_\mu^i) - \tilde\rho_B(\bx,z_\mu^f)\right]
      Y_{lm}^*(\hat{\mbf n})\cr
   & = 1.4 \frac{4 \pi i^{l}}{2 \mu_0\rho_{\gamma0}}
   \int \frac{d^3 k_- d^3 k_1}{(2\pi)^6} 
    \tB(\bk_1) \cdot\tB^*(\bk_2)
    \left[e^{-(k_1^2 + k_2^2)/k_D^2}\right]_f^i
      W\!\left(\frac{k_-}{k_s}\right) j_{l}(k_-r_L) Y_{lm}(\hat\bk_-)\,,\cr
\end{align}
where $\left[f(k_D)\right]_f^i\equiv f(k_D^i) - f(k_D^f)$. We have defined $\bk_-\equiv \bk_1 - \bk_2$ and used the comoving distance to the surface of last scattering $r_L\approx13.9\text{ Gpc}$ (calculated by, e.g. CAMB). Also, we have included a window function $W(k/k_s)\equiv\left.3K^{-3}\big(\sin K-K\cos K\big)\right\rvert_{K=k/k_s}$ (the Fourier transform of the top-hat function), which spreads out the $\mu$-distortion over a physical volume with radius $1/k_s$. This is necessary since the photon distribution at last scattering comes from the combination of various photon distributions within the photon diffusion scale. Indeed we naively expect $k_s$ to be roughly the damping scale at last scattering $k_D^\text{LS}\approx0.14\text{ Mpc}^{-1}$; this is borne out by a more rigorous computation in \cite{Pajer&Zald:2012-A-hydrodynamical}, who find $k_s\approx 0.084\text{ Mpc}^{-1}$ and is the value we will use.

Next, we must correlate (\ref{eq:almu}) with the comparable decomposition of the temperature perturbation, which for adiabatic scalar fluctuations (see, e.g. \cite{Bartolo:2004if}) is 
\begin{align}\label{eq:alT}
  a_{lm}^T = \frac{12\pi}{5} i^l \int \frac{d^3k}{(2\pi)^3} \zeta_\bk g_{Tl}(k) Y_{lm}^*(\hat k)\,,
\end{align}
where we calculate the scalar radiation transfer function $g_{Tl}$ using CAMB \cite{Lewis:1999bs,CAMB}; then, we can find $C_l$
\begin{align}
  \braket{a_{lm}^{*\mu} a_{l'm'}^{T}} = \delta_{ll'} \delta_{mm'} C_l\,.
\end{align}
Correlating (\ref{eq:almu}), (\ref{eq:alT}) and using the rotational symmetry of the $\bbzet$ correlator, we get
\begin{align} \label{eq:ClmuT-simp-expnl}
  C_l^{\mu T}
   & \begin{aligned}[t]
     = 1.4 \frac{6}{5 (2\pi)^3} \frac{1}{\mu_0\rho_{\gamma0}}
      \int dk \,d k_1\, du\, &k^2 k_1^2
       \braket{\tB^*(\bk_1) \cdot\tB(\bk_2) \zeta(\bk)}'
     \left[e^{-(k_1^2 + k_2^2)/k_D^2}\right]_f^i\cr
     &\times W\!\left(\frac{k}{k_s}\right) j_{l}(kr_L) g_{Tl}(k)\,,
   \end{aligned}
\end{align}
where $\bk_2 = \bk_1 - \bk$, and we have defined $\braket{\tB^*(\bk_1) \cdot\tB(\bk_2) \zeta(\bk)} \equiv (2\pi)^3 \,\delta(-\bk_1 + \bk_2 + \bk) \braket{\tB^*(\bk_1) \cdot\tB(\bk_2) \zeta(\bk)}'$.

\subsection{Approximations for simplifying the calculation of $C_l^{\mu T}$}
\label{sec:aprxns}

As written, (\ref{eq:ClmuT-simp-expnl}) is a difficult triple integral. We use some assumptions to simplify it.
\begin{itemize}
\item The window function $W(K) \equiv 3K^{-3}\big(\sin K-K\cos K\big)$ cuts off roughly at $K\approx 1$, i.e. $k \approx k_s$. [Note also the integrand in (\ref{eq:ClmuT-simp-expnl}) has a factor $j_l(k r_L)$ (and we approximately get another such factor from $g_{Tl}(k)\approx \frac{1}{3} j_l(k r_L)$), which already starts cutting off $k$ at $l/r_L \lesssim k_s$)].

Thus, we can cut off the $k$ integral near $k_s$; we will cut it off at $10 k_s$.

\item The exponential factor suppresses $k_1, k_2$ below $k_D$. Also, $k_D^i, k_D^f\gg k_s\gtrsim k$.

Thus, we can assume we are in the squeezed limit, i.e. $k_1\approx k_2\ll k$.

\item We can use the exponential as a cutoff for $k_1, k_2$

Thus,
\begin{align}
  \left.\exp\left(\!- \frac{k_1^2 + k_2^2}{k_{D}^2}\right)\right\rvert_f^i
   \simeq 
   \begin{cases}
     1 & \ck_D^i > k > \ck_D^f\\
     0 & \text{otherwise}
   \end{cases}\,,
\end{align}
where we defined $\ck$ in (\ref{eq:pow-damping-scale}).

\end{itemize}

Then, \eqref{eq:ClmuT-simp-expnl} becomes
\begin{align} \label{eq:ClmuT-simpd}
  C_l^{\mu T}
     \approx 1.4 \frac{6}{5 (2\pi)^3} \frac{1}{\mu_0\rho_{\gamma0}}
      \int_0^{10 k_s} dk \int_{\ck_D^f}^{\ck_D^i} d k_1 \int_{-1}^{1} du \, &k^2 k_1^2 
       \braket{\tB^*(\bk_1) \cdot\tB(\bk_2) \zeta(\bk)}'
        _{k\ll k_1\approx k_2} \cr
     &\times W\!\left(\frac{k}{k_s}\right) j_{l}(kr_L) g_{Tl}(k)\,,
\end{align}
where $\bk_2 = \bk_1 - \bk$. This result lets us calculate $C_l$ quickly once we specify $\braket{\tB^*(\bk_1) \cdot\tB(\bk_2) \zeta(\bk)}$.

\section{The signature from a primordial $b_{\text{NL}}$}
\label{sec:primordial-bnl}

\cite{Jain:2012ga} proposed the parameterization of the $\bbzet$ correlator given in (\ref{eq:dfnn-of-bnl}):
\begin{align}\tag{\ref{eq:dfnn-of-bnl}}
  \left\langle\bB^*(\bk_1) \cdot \bB(\bk_2)\, \zeta(\bk_3) \right\rangle_{k_1\approx k_2\ll k_3} = (2\pi)^3 b_{\text{NL}} 
    \delta^{(3)}\!(-\bk_1 + \bk_2 + \bk_3) P_B(k_1) P_\zeta(k_3)
\end{align}
Inserting this into (\ref{eq:ClmuT-simpd}), we get
\begin{align}\label{eq:ClmuT-from-prim-crltn-W-pow-spr}
  C_{l,b_{\text{NL}}}^{\mu T}
   & \approx 1.4 \frac{12}{5 (2\pi)^3} \frac{1}{\mu_0\rho_{\gamma0}} b_{\text{NL}}
      \int_0^{10 k_s} d k\, k^2 P_\zeta(k) W\!\left(\frac{k}{k_s}\right)
       j_{l}(kr_L) g_{Tl}(k)
     \int_{\ck_D^f}^{\ck_D^i} d k_1\, k_1^2
       P_B(k_1)\,,
\end{align}
which we can be written
\begin{align}\label{eq:ClmuT-in-terms-of-Bmutil}
  C_{l,b_{\text{NL}}}^{\mu T}
    \approx 1.4 \frac{3}{5 \pi} \frac{1}{\mu_0\rho_{\gamma0}} b_{\text{NL}} \bmu^2
     \int_0^{10 k_s} d k\, k^2 P_\zeta(k) W\!\left(\frac{k}{k_s}\right) j_{l}(kr_L) g_{Tl}(k)\,;
\end{align}
we define $\bmu$ as the size of the magnetic fields on $\mu$-distortion scales only, i.e.
\begin{align}\label{eq:dfn-Bmu}
    \bmu^2 \equiv \int_{\ck_D^f}^{\ck_D^i} \frac{d^3 k_1}{(2\pi)^3} P_B(k_1)\,.
\end{align}

We analyze the implications of this result in Sec. \ref{sec:observ-prim-BBzet}.

\section{The signature from the evolution of $\zeta$  due to $B$}
\label{sec:evln-of-zet-due-to-B}

In a pure adiabatic mode, the curvature perturbation $\zeta$ is constant outside the horizon. However, isocurvature modes such as a magnetic field produce anisotropic stress which causes $\zeta$ to evolve, although once neutrinos decouple from baryons and electrons (at around 1 MeV \cite{Weinberg:2008zzc}), they rapidly develop a compensating anisotropic stress that cancels out that from the isocurvature \cite{Lewis:2004ef,Shaw:2009nf}. This evolution produces a correlation between the magnetic field and the curvature perturbation, which could mask a primordial $\bbzet$ correlation. This calculation was
previously done in \cite{Miyamoto:2013oua} and we find the same result as theirs if we make the same assumptions as them\footnote{Actually, their result has a small mistake in equation (31), where they neglect a factor of $3/5$ needed to convert $a^T_{lm}$ in terms of the Newtonian potential factor into the appropriate formula in terms of the curvature perturbation, cf. (\ref{eq:alT}).}.

In principle, magnetic fields also source vector and tensor metric perturbations \cite{Lewis:2004ef,Shaw:2009nf}, which also show up in the CMB \cite{Durrer:1998ya,Mack:2001gc,Lewis:2004ef,Shaw:2009nf,Shiraishi:2010sm,Shiraishi:2011fi,Shiraishi:2012rm,Shiraishi:2012xt}. However, their contribution should be of the same order of magnitude as the scalar contribution, which we find is negligible, so we will not consider these other contributions.

Shaw and Lewis \cite{Shaw:2009nf} find that the contribution to $\zeta$ from a magnetic field can be expressed
\begin{align}
  \zeta_B(\bk) & \approx \frac{3}{2} R_\gamma T_{ij}(\hat k) \Delta^{ij}(\bk) 
      \left[\log\left(\frac{\eta_\nu}{\eta_B}\right) 
       + \left(\frac{5}{8 R_\nu} - 1\right)\right]\,,
\end{align}
where
\begin{align} \label{eq:delta_magdef}
    \Delta^{ij}(\bk) &\equiv \frac{1}{4\pi (2\pi)^3 \tilde\rho_\gamma} 
      \int d^3p\, d^3q\, \tb_i(\mbf p) \tb_j(\mbf q) \delta(\bk - \mbf p - \mbf q) \,,\cr
  T_{ij}(\hat k) &\equiv \hat k_i \hat k_j - \frac{1}{3} \delta_{ij} \,,
\end{align}
\begin{align}
  R_\gamma &\equiv \frac{\rho_\gamma}{\rho_r}\,, & R_\nu &\equiv \frac{\rho_\nu}{\rho_r}\,,
\end{align}
and where $\eta_\nu$ is the time of neutrino decoupling ($T\sim 1$ MeV, \cite{Weinberg:2008zzc}) and $\eta_B$ is the time (during radiation domination) when $\zeta$ begins to evolve under the influence of $B$. If the magnetic field is generated post-inflation, $\eta_B$ is the time of magnetogenesis, whereas if the magnetic field is inflationary, $\eta_B$ should be the end time of reheating, with any effects during inflation included in $b_{\text{NL}}$. Thus, given the results of \cite{Ade:2014xna}, we use $H_B = 10^{14}$ GeV, giving $\eta_\nu/\eta_B\approx 10^{19}$. Actually, finding the exact value of $\eta_B$ is unnecessary, since we take the log.

We then find that 
\begin{align}
  \big\langle \tB^*(\bk_1) \cdot\tB(\bk_2) \zeta_B(\bk) \big\rangle'
    = \frac{1}{2} (2\pi)^3 \xi P_B(k_1) P_B(k_2) g(k_1,k,u) \,,
\end{align}
where we have defined ($u\equiv\hat\bk_1\cdot\bk$),
\begin{align}
  g(k_1,k,u) \equiv
  \frac{k_1^2 \left(1 - 3 u^2\right) - k^2 \left(1+u^2\right) 
      + k_1 k u \left(1+3 u^2\right)}
    {3 (k_1^2 + k^2 - 2 k_1 k u)}\,,
\end{align}
and
\begin{align}
  \xi \equiv \frac{3}{2(2\pi)^3} \frac{R_\gamma}{\mu_0 \rho_\gamma a^4} 
     \left[\log\left(\frac{\eta_\nu}{\eta_B}\right) + \left(\frac{5}{8 R_\nu} - 1\right)\right]
    \approx 4.6\sci{9} \text{ G}^{-2}\,.
\end{align}

Plugging into (\ref{eq:ClmuT-simpd}), we find 
\begin{align}\label{eq:ClmuT-from-evln-zet-W-pow-spr}
  C_{l,\zeta\text{-evolve}}^{\mu T}
   & = 1.4 \frac{3 \xi}{5\mu_0\rho_{\gamma0}} \int_0^{10 k_s} dk \, k^2 \,
       W\!\left(\frac{k}{k_s}\right) j_{l}(kr_L) g_{Tl}(k)
       \int_{\ck_D^f}^{\ck_D^i} d k_1\, k_1^2 P_B^2(k_1)
       \int_{-1}^{1} du\, g(k_1,k,u)\,.
\end{align}
We can considerably simplify this if we remember that the integrand has support where $k_1\approx k_2\ll k$, so that we can expand the integral $u$ to lowest order in $k/k_1$, i.e.
\begin{align}
    \int_{-1}^1 du\,g(k_1,k,u) \approx - \frac{16}{45} \frac{k^2}{k_1^2}\,.
\end{align}
Then,
\begin{align}
  C_{l,\zeta\text{-evolve}}^{\mu T}
   & \approx  - 1.4 \frac{16 \xi}{75 \mu_0\rho_{\gamma0}} 
      \int_0^{10 k_s} d k\, k^4 
       W\!\left(\frac{k}{k_s}\right)
       j_{l}(kr_L) g_{Tl}(k)
      \int^{\ck_D^i}_{\ck_D^f} d k_1 \, 
       P_B^2(k_1)
\end{align}
where the integral over $k$ must be done numerically, and the integral over $k_1$ must wait until we specify $P_B$. 

\section{The signature from the evolution of $B$ due to $\zeta$}
\label{sec:evln-of-B-due-to-zet}

We next consider the reverse of the effect in the previous section, that is, the evolution of $B$ under the influence of a primordial curvature perturbation, as hypothesized in \cite{Kunze:2012fd}. Naively, one would not expect the evolution of the universe to induce correlations between the widely different $B$ and $\zeta$ scales (and, indeed, we ultimately find little effect). For metric perturbations, we will use the notation of Kodama \& Sasaki \cite{Kodama:1985bj} with one exception (used fairly frequently, e.g. \cite{Shaw:2009nf}): what Kodama \& Sasaki \cite{Kodama:1985bj} call $\Delta_s$, i.e. the gauge-invariant perturbation equal to $\delta$ in Newtonian gauge, we will simply call $\Delta$. 

Following \cite{Kunze:2012fd}, we note that one of Maxwell's equations (keeping quantities up to their second order in perturbations) is \cite{Durrer:2013pga}
\begin{align}\label{eq:Maxwell-eqn-in-crvd-spc}
  \tB' = \mbf{\nabla}\times\left[\mbf{v}_b\times\tB\right] 
   = \left(\tB\cdot\nabla\right) \mbf v_b - \left(\mbf v_b\cdot\nabla\right) \tB
      - \tB \left(\nabla\cdot\mbf v_b\right)\,,
\end{align}
where $\mbf v_b$ is the baryon velocity. We can decompose $\mbf v_b$ into a scalar and vector part
\begin{align}
  \mbf v_b = \boldsymbol{\nabla} v_b^{(0)} + \mbf v_b^{(1)},
\end{align}
where $\nabla\cdot\mbf v_b^{(1)} = 0$. Any primordial vector $v_b^{(1)}$ would redshift away rapidly in an FRW universe, and we neglect it in favor of the scalar part. Also, we drop the $b$ subscript on $v_b$ since in the tight-coupled plasma, all velocities are the same, as well as the $0$ subscript. Then, \eqref{eq:Maxwell-eqn-in-crvd-spc} becomes
\newcommand{\bq}{\mathbf{q}}
\begin{align}\label{eq:B-from-zet-evltn-eqn}
  \tb_i'(\bk,t)
   &\approx \int \frac{d^3q}{(2\pi)^3} \frac{q_i q_j - \big(\bk \cdot \bq\big) \delta_{ij}}{q}
      \tb_j(\bk - \bq,t) v(\bq,t)\,.
\end{align}

Linear FRW perturbations obey (see, e.g. \cite{Doran:2003xq})
\begin{align}
  \Delta' + 3 (c_s^2 - w) \mcl H \Delta + k (1+w) V + 3\mcl H w \Gamma = 0\,.
\end{align}
For an adiabatic perturbation, $\Gamma = 0$, and for radiation domination, $c_s^2 - w = 0$. Thus, in Newtonian gauge, where $\Delta=\delta$ and $V=v$, we find
\begin{align}
  \delta' + \frac{4}{3} k v = 0\,,
\end{align}
which turns (\ref{eq:B-from-zet-evltn-eqn}) into
\begin{align}
  \tb_i'(\bk,t)
   &\approx - \frac{3}{4} 
     \int \frac{d^3q}{(2\pi)^3} \frac{q_i q_j - \big(\bk \cdot \bq\big) \delta_{ij}}{q^2}
      \tb_j(\bk - \bq,t) \delta'(\bq,t)\,.
\end{align}
Since we expect this effect to be small, let us split the magnetic field into  $\tB(\bk,t) = \tB^{(i)}(\bk) + \tilde {\mbf b}(\bk,t)$, where $b\ll B$ and the evolution due to $\delta$ is contained entirely within $\tilde {\mbf b}$. We thus get\footnote{Note that this differs from \cite{Kunze:2012fd}, which has no $\Delta$'s, an important distinction.}
\begin{align}
  \Delta\tb_i(\bk) = \tilde b_i(\bk)
   &\approx - \frac{3}{4} 
     \int \frac{d^3q}{(2\pi)^3} \frac{q_i q_j - \big(\bk \cdot \bq\big) \delta_{ij}}{q^2}
      \tb_j^{(i)}(\bk - \bq,t) \Delta \delta(\bq)\,,
\end{align}
where $\Delta X$ is the change in $X$ during radiation domination. 

However, in Newtonian gauge, $\delta$ is constant outside the horizon, and $\Delta\delta(\bk)\approx0$ for superhorizon $k$. But it is precisely these superhorizon modes that are relevant for calculating $\bbzet\sim \braket{B^{(i)} B^{(i)}} \braket{\zeta\,\Delta\delta}$ from the evolution of $B$. To see this, note that the curvature perturbation $\zeta$ in the correlation $\bbzet$ is observed via the CMB temperature perturbation and is sensitive only to CMB scales, which are outside the horizon during the $\mu$-distortion era. But we correlate $\Delta\delta$ with $\zeta$ in $\braket{\zeta\,\Delta\delta}$, so that $\Delta\delta$ is only sensitive to the same superhorizon scales.

We conclude therefore that there is no $\bbzet$ contribution from the evolution of $B$. (This result is in agreement with the graphs in \cite{Kunze:2012fd}, which show that the $C_l$'s from this effect is considerably too small to be detected in the CMB.)

\section{The observability of primordial $\bbzet$}
\label{sec:observ-prim-BBzet}

\subsection{Parameterizing $B$, $\zeta$}
\label{sec:pow-spr}

For $\zeta$, we use the standard power spectrum with a constant $n_s$
\begin{align}
  P_\zeta(k) &= \frac{(2\pi)^3}{4\pi} \Delta^2(k_p) \frac{1}{k^3} \left(\frac{k_p}{k}\right)^{1-n_s} 
    \,,
\end{align}
and use values from \cite{Planck-overview}, i.e. $k_p=0.05\text{ Mpc}^{-1}$, $n_s=0.9611$, and $\Delta^2 = 2.214\sci{-9}$.

For magnetic fields, only the modes on $\mu$-distortion scales are relevant, so we will write results in term of the field strength $\bmu$ on $\mu$-scales defined in Eq. \eqref{eq:dfn-Bmu}. It is sometimes also necessary to specify an explicit spectrum, which we assume to be of the form
\begin{align}
  P_B(k) = A k^{n_B}\,,
\end{align}
so that $n_B=-3$ is scale-invariant; we can then use \eqref{eq:dfn-Bmu} to derive $A$, giving
\begin{align}
  P_B(k) &
    = 2 \pi^2 \tb_\mu^2 (3+n_B) 
     \frac{k^{n_B}}{k_{\check i}^{3+n_B} - k_{\check f}^{3+n_B}}\,.
\end{align}
One generically imagines that inflation would give $n_B\approx -3$ (scale-invariant), while phase transitions, e.g., often give $n_B=2$ \cite{Durrer:2013pga}. However, our main result is not dependent on $n_B$.

\subsection{Calculating signal-to-noise}
\label{sec:sign-noise}

An observable is detectable when the signal-to-noise becomes order unity. Calculating the signal-to-noise for the $\braket{\mu T}$ correlation is straightforward \cite{Pajer&Zald:2012-New-window,Ganc:2012ae}:
\begin{align}
  \left(\frac{S}{N} \right)^2
   &= \sum_l \frac{\big(C^{\mu T}_l\big)^2}{\sigma_l^2}\,,
\end{align}
where $\sigma_l$ is the variance at multipole $l$ and can be approximated as
\begin{align}
  \sigma_l^2 = \frac{1}{2l+1} \left(C_l^{TT} + C_l^{TT,N}\right)
    \left(C_l^{\mu\mu} + C_l^{\mu\mu,N}\right)
   \approx \frac{1}{2l+1} C_l^{TT} \left(C_l^{\mu\mu} + C_l^{\mu\mu,N}\right)\,,
\end{align}
and where $C_l^{XX,N}$ is the noise in the measurement of a particular correlation. We can neglect $C_l^{TT,N}$ since the $TT$ correlation is already signal dominated. 

The noise in the $\mu\mu$ correlation is instrument dependent and given by \cite{Pajer&Zald:2012-New-window,Ganc:2012ae}
\begin{align}
  C_{l}^{\mu\mu,\text{N}} = N_\mu\exp\left(\frac{l^2}{l_\text{max}^2}\right)\,,
\end{align}
where $l_\text{max}$ is given -- in terms of the instrument's full-width-at-half-max beam size $\theta_b$ -- by $l_\text{max} = \frac{8\ln 2}{\theta_b^2}$, and $N_\mu$ is the instrument's noise level. The proposed satellite experiment PIXIE \cite{Kogut:2011xw} has
\begin{align}\label{eq:PIXIE-parms}
  N_\mu^\text{PIXIE} &= 4\pi\times 10^{-16}\,, & l_{\text{max}}^\text{PIXIE} &\approx 84\,,
\end{align}
which we will often use as a standard figure of merit for determining the detectability of our results.

PIXIE is capable of making an absolute measure of the CMB spectrum. However, this capability is not actually necessary to measure anisotropic $\mu$-distortion, as noted by \cite{Ganc:2012ae}. Anisotropic $\mu$-distortion can in principle be probed by relatively calibrated experiment like Planck. Planck, for example, might be expected to have parameters $N_\mu^\text{Planck} \approx 10^{-15}$, $l_{\text{max}}^\text{Planck} \approx 861$, and thus be comparable in sensitivity to PIXIE. (The difference in $l_\text{max}$ is less significant than it appears since, as noted by \cite{Pajer&Zald:2012-New-window}, the signal goes roughly as $\sqrt{\ln (l_\text{max})/2}$.) There is, however, a potential calibration issue with relatively calibrated experiment, since they involve comparing the response of two different frequencies; this effect must be properly considered if attempting such a measurement and could degrade the instrument's performance beyond what we calculate here.

We will also use figures from the proposed CMBPol \cite{Baumann:2008aq} satellite, a relatively calibrated instrument like Planck but with a better noise response. The parameters for CMBPol \cite{Miyamoto:2013oua} are
\begin{align}\label{eq:CMBPol-parms}
  N_\mu^\text{CMBPol} &\approx 2\sci{-18}\,, & l_{\text{max}}^\text{CMBPol} &\approx 1000\,,
\end{align}
an improvement over PIXIE though with the caveat above about relatively calibrated experiments.

We can calculate $\Braket{\mu\mu}$ similarly to how we calculated $\Braket{\mu T}$, beginning with the auto-correlation of (\ref{eq:almu}). Using the same approximations described in Sec. \ref{sec:aprxns}, we find that
\begin{align}
  C_l^{\mu\mu}
   & \approx (1.4)^2 \frac{2}{(2\pi)^3} \frac{1}{\mu_0^2 \rho_{\gamma0}^2}
     \int_{\ck_D^f}^{\ck_D^i} dk_1\, k_1^2 P_B^2(k_1)
      \int_0^{10 k_s} dk_-\,  k_-^2
       \left[W\!\left(\frac{k_-}{k_s}\right)\right]^2 j_l^2(k_-r_L)\,.
\end{align}
Plotting $C_l^{\mu\mu}$ from this formula matches the graphs in \cite{Miyamoto:2013oua} for the same parameters for $l<1000$. However, to get accurate results, one needs to consider the window function, particularly for $l\gtrsim k_s r_L\approx 1200$.

We find (as did \cite{Miyamoto:2013oua}) that, for the cases relevant here and for instruments in the foreseeable future, the noise term for $\braket{\mu\mu}$ correlations dominates over the signal. Thus, our formulas neglect $C_l^{\mu\mu}$.

\subsection{The signal from a primordial $b_{\text{NL}}$}

\begin{figure}\label{fig:ClmuT-prim}
  \centering
  \includegraphics{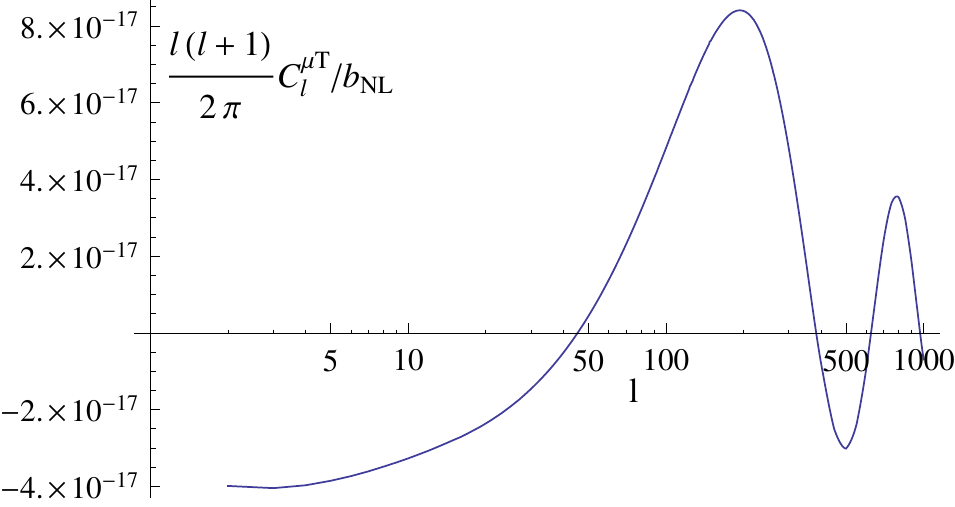}
  \caption{The $C_{l,b_{\text{NL}}}^{\mu T}$ spectrum from a primordial $\bbzet$ correlation parameterized by $b_{\text{NL}}$. We show the result for $\bmu=1$ nG. The oscillations come from baryon acoustic oscillations.}
\end{figure}

Inserting the power spectrum for $\zeta$ into formula \eqref{eq:ClmuT-in-terms-of-Bmutil} for $C_l^{\mu T}$, we arrive at
\begin{align}\label{eq:ClmuT-from-prim-crltn-substd}
  C_l^{\mu T}
   & \approx \left(1.2\sci{26}\text{ G}^{-2}\text{\;Mpc}^{-2(2-n_s)}\right)\times
     b_{\text{NL}} \bmu^2
      \int_0^{10 k_s} d k\, k^{-(3-n_s)} j_1\left(\frac{k}{k_s}\right) 
         j_{l}(kr_L) g_{Tl}(k)\,.
\end{align}
where the integral must be numerically evaluated. We plot the shape of the spectrum in Fig. \ref{fig:ClmuT-prim}.

For a separable parameterization of the correlation $\bbzet$, as in our parameterization of $b_{\text{NL}}$ in (\ref{eq:dfnn-of-bnl}), there is no spectral shape dependence on the details of the magnetic field spectrum, only on the net strength $\bmu$. If, however, we had a scale-dependent $b_{\text{NL}}$, this could alter the shape of the spectrum. We will not pursue this line of inquiry here, though it might be interesting given a well-motivated model.

We can proceed to calculate the signal-to-noise for the PIXIE and CMBPol experiments:
\begin{align}\label{eq:SToN-bNL}
  \frac{S}{N}
    = b_{\text{NL}} \left(\frac{\bmu}{1\text{ nG}}\right)^2 \times
    \begin{cases}
      8.4\sci{-4} & \text{PIXIE} \\
      2.4\sci{-2} & \text{CMBPol}
    \end{cases} \,.
\end{align}
We thus see that $\mu$-distortion is sensitive to the combination $b_{\text{NL}} \left(\frac{\bmu}{1\text{ nG}}\right)^2$, i.e. if $b_{\text{NL}}=\sO(1)$, we would need $\bmu\approx 10$ nG to see the result in CMBPol (though if $b_{\text{NL}}$ were larger, $\bmu$ could be smaller). Thus, $\mu$-distortion can provide meaningful constraints on a primordial $b_{\text{NL}}$ correlation. 

We can bound the magnetic fields on $\mu$-distortion scales through the sky-averaged $\mu$ monopole. From \eqref{eq:mu-distortion}, we find that $\mu$-monopole generated by the damping of a magnetic field $\bmu$ is
\begin{align}
  \mu \approx 1.4 \frac{\rho_{\bmu}}{\rho_\gamma} 
   = 1.4 \frac{\bmu^2/2\mu_0}{\rho_{\gamma0}}\,.
\end{align}
Thus, the limits from COBE/FIRAS+TRIS ($\mu<6\sci{-5}$), PIXIE ($\mu<9\sci{-8}$) constrain
\begin{align}
  \bmu< \sqrt{\frac{2 \mu \mu_0 \rho_{\gamma0}}{1.4}} = \sqrt\mu \left(3.6\sci{-6}\text{ G}\right) =
  \begin{cases}
    27\text{ nG}& \text{FIRAS}\\
    1\text{ nG}& \text{PIXIE}\\
  \end{cases}
\,.
\end{align}
(Note that this result agrees with Fig. 1 of \cite{Kunze:2013uja}, where they more thoroughly investigated the signal of the $\mu$-monopole from magnetic fields). Thus, we already know that $\bmu<27 nG$. If the primordial magnetic fields saturated this limit, PIXIE would detect $\Braket{\mu T}$ for $b_{\text{NL}}=\sO(1)$.

On the other hand, considerations of inflationary magnetogenesis lead us to think there will not be a signal. Inflation provides a potential mechanism for producing large-scale magnetic fields that are naturally correlated with the curvature perturbation. However, the strength of inflationary magnetic fields is quite limited in high scale inflation \cite{Ferreira:2013sqa,Fujita:2014sna,Ferreira:2014af}; for magnetic fields produced during inflation by a gauge-invariant coupling that avoids both the strong coupling and backreaction problems, the model-independent bound (\cite{Fujita:2014sna}) is
\begin{equation}
  \tilde B < \left( 5\sci{-14}\text{ nG} \right)
   \left(\frac{\rho_\text{inf}^{1/4}}{10^{14}\text{ GeV}}\right)^{-1}    
   \left(\frac{k_B}{\ck_D^i}\right)^{\frac{5}{4}} \,,
\end{equation}
where $k_B$ indicates the wavenumber where the magnetic field's power is concentrated.
If one takes the recent BICEP2 results \cite{Ade:2014xna} at face value, then $\rho_\text{inf}^{1/4}\approx 10^{14}$ GeV, leading to the constraint that, on $\mu$-distortion scales, $\tilde B< 5\sci{-16}$ nG, giving a minuscule signal-to-noise. Thus, if the BICEP2 result holds up, a detection of primordial magnetic field correlation from inflation would seem to imply that magnetic fields are generated in a regime of either strong coupling, backreaction or broken gauge invariance\footnote{Of course, if magnetic fields themselves are responsible for the B-mode signal in BICEP2, as speculated in \cite{Bonvin:2014xia}, then none of these conclusions apply.}, which seems discouraging from a theoretical point of view. (But on the other hand, if the detection of intercluster magnetic fields -- mentioned in the introduction -- is validated, we must still find some mechanism for their production.)

\subsection{A competing signal: $\braket{\mu T}$ correlation from a primordial bispectrum}
\label{sec:signal-from-fNL}

\cite{Pajer&Zald:2012-A-hydrodynamical} originally proposed anisotropic $\mu$-distortion as a means of measuring the squeezed limit bispectrum of the density perturbation. Therefore, such an effect can compete with a $\braket{\mu T}$ signal from a $\bbzet$ correlation. The signal from $f_{\text{NL}}$ is generated by diffusion damping of density perturbations during the $\mu$-era, giving a $\mu$-distortion as per Eq. (\ref{eq:mu-distortion}); the energy here comes from the damping of density waves with energy density (\cite{Pajer&Zald:2012-A-hydrodynamical}) $\rho = \rho \braket{\delta^2}_p \frac{c_s^2}{1 + w} = \frac{1}{4} \rho \braket{\delta^2}_p$, where $\braket{{}}_p$ indicates an average over the wave's oscillation period. To connect with the primordial perturbation, we need to know that, for $k$ deep within the horizon during radiation domination, $\delta_\bk \approx -4\zeta_\bk \exp(-k^2/k_{D\gamma}^2)\cos\big(kr(t)\big)$, where $k_{d\gamma}$ is the photon diffusion scale in (\ref{eq:photon-diff-scale}) and $r(t) \approx 2 t/a\sqrt{3}$. Proceeding in a similar manner as in Sec. \ref{sec:prod-mu-dist} (and using the same assumptions), we find
\begin{align}\label{eq:ClmuT-from-dnsty-prtbns}
  C_{l,f_{\text{NL}}}^{\mu T}
   \approx 1.4 \frac{576}{25 (2 \pi)^3} f_{\text{NL}}
    \int_0^{10 k_s} dk\, k^2 P_\zeta(k) W\!\left(\frac{k}{k_s}\right) j_{l}(k r_L) g_{Tl}(k) 
     \int_{\ck_{D\rho}^f}^{\ck_{D\rho}^i} dk_1 \, k_1^2 P_\zeta(k_1)\,.
\end{align}
In principle, our knowledge of the power spectrum for the $k$'s relevant to $\mu$-distortion is very limited but for simplicity here we will simply assume the same power spectrum throughout. (This is particularly questionable if the tension between Planck \cite{Planck-overview} and BICEP2 is resolved by a negative running of the spectral index, which could suppress the bispectrum signal relative to the magnetic field signal).

By comparing \eqref{eq:ClmuT-from-prim-crltn-substd} and \eqref{eq:ClmuT-from-dnsty-prtbns} we see that the spectral shape of the $\braket{\mu T}$ is the same from both damping of density perturbations and damping of magnetic fields. Thus, unfortunately, studying the correlation spectrum will not help us distinguish between the source of a detection. There is a possible exception if either $f_{\text{NL}}$ and $b_{\text{NL}}$ deviate from scale-invariance\footnote{Measuring an unusual scaling of $f_{\text{NL}}$ in $\mu$-distortion was the original motivation for \cite{Ganc:2012ae}, which was motivated by earlier work in \cite{Agullo:2010ws,Ganc:2011dy}.} but it would be very difficult to reliably presuppose a particular scaling of either $f_{\text{NL}}$ or $b_{\text{NL}}$ in order to interpret data.

We can try to get some guidance by predicting the size of $f_{\text{NL}}$. As a lower bound, in single-field inflation, one generically expects $f_{\text{NL}}=\frac{5}{12} (1-n_s)\approx 0.01$ (\cite{Maldacena:2002vr,Creminelli:2004yq}) (although there are potential exceptions, e.g. \cite{Ganc:2010ff,Agullo:2010ws,Ganc:2011dy,Namjoo:2012aa,Chen:2013aj,Chen:2013eea}). This expression gives the result
\begin{align}
  \left(\frac{S}{N}\right)_{f_{\text{NL}}} =
   \big\lvert f_{\text{NL}}\big\rvert
  \begin{cases}
    4\sci{-4} & \text{for PIXIE}\\
    1\sci{-2} & \text{for CMBPol}
  \end{cases}\,.
\end{align}
Thus, the generic single field inflation estimate for $f_{NL}$ is not likely to produce a measurable $\braket{\mu T}$ correlation in the foreseeable future. 

On the other hand, the magnetic field itself will also source non-Gaussianity during inflation \cite{Caprini:2009vk,Barnaby:2012tk,Lyth:2013sha,Nurmi:2013gpa}. Since we do not have a single preferred model of inflationary magnetogenesis, we can try to make a fairly model-independent estimate of this effect. If the spectrum of magnetic fields during inflation is approximately Gaussian and has the simple power law form $P_B(k)=C_B k^{-3}\left(k/aH\right)^{\alpha}$, the non-adiabatic pressure of the magnetic field induces a curvature perturbation $\zeta_B= -\int dt H \delta P_\text{nad} /(\rho+p)$, where $\delta P_\text{nad} =(4/3) \rho_B$ is proportional to $P_B$; $\zeta_B$, then, is very non-Gaussian since it is proportional to $B^2$. Using this simple relation, we can easily calculate a lower bound on the non-Gaussianity induced during inflation by the magnetic field.

In the scale invariant case, the dominant contribution in the squeezed limit is from $\left<\zeta_B(\bk_1)\zeta_B(\bk_2)\zeta_B(\bk_3)\right>$. However, to produce a $\langle\mu T\rangle$ signal, there must be inflationary magnetic fields on the larger CMB scale $k_3$. Given the scope of this paper, we will simply assume this is not the case and that there is thus no $\left<\zeta_B^3\right>$ contribution. [If desired, this correlation could be calculated in a straightforward fashion from, e.g., \cite{Nurmi:2013gpa}]. On the other hand, a correlation of the form $\left<\zeta_0(\bk_1)\zeta_B(\bk_2)\zeta_B(\bk_3)\right>$ is unavoidable if $b_{NL}$ is non-zero, where $\zeta_0$ is the curvature perturbation from the inflaton at horizon crossing. Indeed, defining the strength of the magnetic field today on the scale $1/k$ as $B^2_0(k)/2 \equiv d\rho_{B_0}(k)/d\ln(k)$, we find from \cite{Nurmi:2013gpa} 
\begin{equation}
f_{NL} \gtrsim 0.2 \alpha^{-3} \left(\frac{B_0(k)}{1 \textrm{nG}}\right)^4 \left(\frac{0.01}{\epsilon}\right)^2\left(\frac{1-e^{-\alpha N_k}}{e^{-\alpha N_k}}\right)^2 e^{-\alpha (N_0-N_k)}
\end{equation}
which in the exactly scale invariant case $\alpha=0$ becomes
\begin{equation}
f_{NL} \gtrsim 0.2  \left(\frac{B_0(k)}{1 \textrm{nG}}\right)^4 \left(\frac{0.01}{\epsilon}\right)^2 N_k^2(N_0-N_k)~.
\end{equation}
Here $N_k$ is the number of e-folds before the end of inflation and when the relevant scale left the horizon, and $N_0$ is the number of e-folds before the end of inflation and when the generation of magnetic fields started, i.e. it is the infrared cutoff on the power spectrum of magnetic fields during inflation.

Identifying $B_0(k) \approx \tilde B_{\mu}$ implies that the induced $f_{\text{NL}}\propto b_{\text{NL}} \bmu^4$ will be greater than $b_{\text{NL}} \big({\bmu}/{1\text{ nG}}\big)^2$ for $\bmu\gtrsim 1$ nG, and therefore, in this regime, the signal induced by $b_{\text{NL}}$ in the damping of density perturbations will exceed that from the damping of magnetic fields. Notably, though, upcoming experiments are only sensitive to $b_{\text{NL}} (\bmu/1\text{ nG})^2 \gtrsim 100$ so that, unless $b_{\text{NL}}$ is large, a magnetic field signal in $\mu$-distortion would be mostly apparent through $f_{\text{NL}}$.

Note that this derivation assumes that the magnetic field energy was produced during inflation and decayed afterwards at the same rate as the rest of the universe. If on the other hand the inflationary magnetic fields were further enhanced after the end of inflation (as for example if the universe went through a period where it was a stiff fluid (with $w=1$) \cite{Ferreira:2013sqa,Demozzi:2012wh,Martin:2007ue}, or as in the scenario of \cite{Kobayashi:2014ag}), then the constraints of \cite{Nurmi:2013gpa} do not apply directly. But as long as the magnetic fields are enhanced before the $\mu$-distortion era, the result in (\ref{eq:SToN-bNL}) still applies.  

\subsection{The signal from evolution of $\zeta$}
\label{sec:sign-from-evvn-of-zet}

We now briefly discuss the signal from the evolution of $\zeta$ described in Sec. \ref{sec:evln-of-zet-due-to-B}, though we will see that it is very small.

Substituting the power spectrum into \eqref{eq:ClmuT-from-evln-zet-W-pow-spr}, we get
\begin{align}
  C_{l,\zeta\text{-evolve}}^{\mu T}
   \approx &- \left(2.6\sci{22}\text{ G}^{-4}\;\text{Mpc}^{-1}\right)\times \bmu^4 
    \frac{(3+n_B)^2}{1+2n_B}
    \frac{k_{\check i}^{1+2n_B} - k_{\check f}^{1+2n_B}}
      {\left(k_{\check i}^{3+n_B} - k_{\check f}^{3+n_B}\right)^{2}} \cr
    &\times\int_0^{10 k_s} d k\, k^3 j_1\left(\frac{k}{k_s}\right) j_{l}(kr_L) g_{Tl}(k) \,.
\end{align}
We find that
\begin{align}
  \left(\frac{S}{N}\right)_{\zeta\text{-evolve}}
   &= \bmu^4
    \frac{(3+n_B)^2}{1+2n_B}
    \frac{k_{\check i}^{1+2n_B} - k_{\check f}^{1+2n_B}}
     {\left(k_{\check i}^{3+n_B} - k_{\check f}^{3+n_B}\right)^{2}}
     \times
     \begin{cases}
       4.5\sci{36} \text{ G}^{-4}\;\text{Mpc}^{-5} & \text{PIXIE}\\
       1.1\sci{41} \text{ G}^{-4}\;\text{Mpc}^{-5} & \text{CMBPol}
     \end{cases}\,,
\end{align}
or, for a scale-invariant spectrum $n_B=-3$,
\begin{align}
  \left(\frac{S}{N}\right)_{\zeta\text{-evolve}}  
   = \left(\frac{\bmu}{1\text{ nG}}\right)^4 \times
   \begin{cases}
     4.2\sci{-23} & \text{PIXIE}\\
     1.1\sci{-18} & \text{CMBPol}
   \end{cases}\,.
\end{align}
Not only is this effect small, it is likely far smaller than the effect from $f_{\text{NL}}$ (even from just the single field inflation contribution), so it is hard to imagine that it could be measured.

\section{Conclusion}
\label{sec:conclusion}

In this work, we have investigated the sensitivity of the $\braket{\mu T}$ signal to a primordial $\bbzet$ correlation, analyzing the signal in upcoming experiments as well as considering the other possibly competing magnetic and density perturbation effects. We found that a CMBPol-like experiment can constrain $b_{\text{NL}} (\bmu/1\text{ nG})^2 \lesssim 100$, where $\bmu$ is the magnetic field on $\mu$-distortion scales. Thus, the $\braket{\mu T}$ correlation could meaningfully constrain primordial magnetic field correlations.

We can also say with certainty that we will not have measurable post-magnetogenesis contributions to the $\bbzet$ correlator. The signal from the evolution of $\zeta$ due to magnetic fields, considered in Secs. \ref{sec:evln-of-zet-due-to-B}, \ref{sec:sign-from-evvn-of-zet} and previously in \cite{Miyamoto:2013oua}, is too small to affect a $b_{\text{NL}}$ measurement. We also saw in Sec. \ref{sec:evln-of-B-due-to-zet} that, due to the lack of superhorizon velocity flows in the early universe, we do not have a correlation induced by the evolution of $B$ due to initial perturbations. The small size of these effects is, perhaps, unsurprising because it is generally difficult to generate correlations between very different length scales.

The signal from a squeezed-limit three-point function $f_{\text{NL}}$, the initial motivation for anisotropic $\mu$-distortion in \cite{Pajer&Zald:2012-New-window}, can potentially compete with a $b_{\text{NL}}$ signal, depending on the relative strengths of $f_{\text{NL}}$ and $b_{\text{NL}}$ and the size of the magnetic field and curvature perturbation on $\mu$-distortion scales. The shape of $C_l^{\mu T}$ is the same in both cases if both $f_{\text{NL}}$ and $b_{\text{NL}}$ are scale-invariant, so shape information cannot easily help us identify the source of a signal unless we have a motivated theoretical prior. In principle, one may be able to measure all of the $\mu$ monopole, $\braket{\mu T}$, and $\braket{\mu\mu}$ correlations; combining them, it might be possible to say more about the sources of a signal.

We cannot neglect another important consideration: a magnetic field produced during inflation would, through its non-adiabatic pressure over the evolution of the universe, induce an $f_{\text{NL}}$. We discussed in Sec. \ref{sec:signal-from-fNL} that generically, for $\bmu\gtrsim 1$ nG, the induced $f_{\text{NL}}$ signal would dominate over the $b_{\text{NL}}$ signal from magnetic field damping. Since upcoming experiments will not be sensitive to such magnetic fields strengths unless $b_{\text{NL}}$ is large, the $f_{\text{NL}}$ signal induced from $b_{\text{NL}}$ would be more relevant for constraining magnetic fields in the near future. Note that this assumes the magnetic field is produced during inflation and then decays at the same rate as the rest of the universe.

There are at least a few issues worth considering further. In particular, if one is interested in connecting the magnetic field strengths in the $\mu$-era to even earlier times, e.g. magnetogenesis, one should carefully explore the typically non-trivial evolution of magnetic fields in the early universe. It would also be interesting to consider if there are any other mechanisms that could produce a $\braket{\mu T}$ correlation besides a primordial $\bbzet$ correlation or density perturbation bispectrum.

\acknowledgments

We would like to thank Eiichiro Komatsu for early discussions about $\mu$-distortion from magnetic fields, Kerstin Kunze for explaining aspects of her paper about the evolution of $\zeta$, and Rishi Khatri for discussing some aspects of $\mu$-distortion. MSS is supported by a Jr. Group Leader Fellowship from the Lundbeck Foundation. The CP$^3$-Origins centre is partially funded by the Danish National Research Foundation, grant number DNRF90.

\providecommand{\href}[2]{#2}\begingroup\raggedright\endgroup

\end{document}